\documentclass[a4]{article}

\usepackage{epsf}

\DeclareSymbolFont{boldletters}{OML}{cmm}{b}{it}
\DeclareSymbolFontAlphabet{\mathbit}{boldletters}
\DeclareMathSymbol{\alpha}{\mathalpha}{letters}{"0B}
\DeclareMathSymbol{\beta}{\mathalpha}{letters}{"0C}
\DeclareMathSymbol{\gamma}{\mathalpha}{letters}{"0D}
\DeclareMathSymbol{\delta}{\mathalpha}{letters}{"0E}
\DeclareMathSymbol{\epsilon}{\mathalpha}{letters}{"0F}
\DeclareMathSymbol{\zeta}{\mathalpha}{letters}{"10}
\DeclareMathSymbol{\eta}{\mathalpha}{letters}{"11}
\DeclareMathSymbol{\theta}{\mathalpha}{letters}{"12}
\DeclareMathSymbol{\iota}{\mathalpha}{letters}{"13}
\DeclareMathSymbol{\kappa}{\mathalpha}{letters}{"14}
\DeclareMathSymbol{\lambda}{\mathalpha}{letters}{"15}
\DeclareMathSymbol{\mu}{\mathalpha}{letters}{"16}
\DeclareMathSymbol{\nu}{\mathalpha}{letters}{"17}
\DeclareMathSymbol{\xi}{\mathalpha}{letters}{"18}
\DeclareMathSymbol{\pi}{\mathalpha}{letters}{"19}
\DeclareMathSymbol{\rho}{\mathalpha}{letters}{"1A}
\DeclareMathSymbol{\sigma}{\mathalpha}{letters}{"1B}
\DeclareMathSymbol{\tau}{\mathalpha}{letters}{"1C}
\DeclareMathSymbol{\upsilon}{\mathalpha}{letters}{"1D}
\DeclareMathSymbol{\phi}{\mathalpha}{letters}{"1E}
\DeclareMathSymbol{\chi}{\mathalpha}{letters}{"1F}
\DeclareMathSymbol{\psi}{\mathalpha}{letters}{"20}
\DeclareMathSymbol{\omega}{\mathalpha}{letters}{"21}
\DeclareMathSymbol{\varepsilon}{\mathalpha}{letters}{"22}
\DeclareMathSymbol{\vartheta}{\mathalpha}{letters}{"23}
\DeclareMathSymbol{\varpi}{\mathalpha}{letters}{"24}
\DeclareMathSymbol{\varrho}{\mathalpha}{letters}{"25}
\DeclareMathSymbol{\varsigma}{\mathalpha}{letters}{"26}
\DeclareMathSymbol{\varphi}{\mathalpha}{letters}{"27}
\DeclareMathSymbol{\Gamma}{\mathalpha}{letters}{"00}
\DeclareMathSymbol{\Delta}{\mathalpha}{letters}{"01}
\DeclareMathSymbol{\Theta}{\mathalpha}{letters}{"02}
\DeclareMathSymbol{\Lambda}{\mathalpha}{letters}{"03}
\DeclareMathSymbol{\Xi}{\mathalpha}{letters}{"04}
\DeclareMathSymbol{\Pi}{\mathalpha}{letters}{"05}
\DeclareMathSymbol{\Sigma}{\mathalpha}{letters}{"06}
\DeclareMathSymbol{\Upsilon}{\mathalpha}{letters}{"07}
\DeclareMathSymbol{\Phi}{\mathalpha}{letters}{"08}
\DeclareMathSymbol{\Psi}{\mathalpha}{letters}{"09}
\DeclareMathSymbol{\Omega}{\mathalpha}{letters}{"0A}

\newcommand{\mbit}[1]{{\mathbit#1}}

\begin{document}

\title{\begin{flushright}
\small
Ehime-th-1 \\
Kyushu-HET-64 \\
\end{flushright}
Caustics in the Grassmann Integral}
\author{Taro KASHIWA\thanks{kashiwa@phys.sci.ehime-u.ac.jp}\\
Department of Physics, Ehime University, Matsuyama, 790-8577, Japan\\ and 
\\
Tomohiko SAKAGUCHI\thanks{tomohiko@higgs.phys.kyushu-u.ac.jp}\\
Department of Physics, Kyushu University, Fukuoka, 812-8581, Japan}
\date{\today}

\maketitle

\abstract{It is shown that a simple model of $2N$-Grassmann variables  with a four-body coupling involves caustics when the integral has been converted to a bosonic form with the aid of the auxiliary field. Approximation is then performed to assure validity of the auxiliary field method(AFM). It turns out that even in $N=2$, the smallest case in which a four-body interaction exists, AFM does work more excellently if higher order effects, given by a series in terms of $1/ \   {}^3 \!  \! \! \! \sqrt{N}$ around a caustic and of $1/N$ around a saddle point, would be taken into account.}

\newpage

\section{Introduction}
In the path integral formalism, it is well-known that the semiclassical or the WKB approximation is a straightforward and moreover the most plausible prescription in nonpertubative treatments\cite{rf:KOS}. Under that, first we must find out solution(s) of the equation deriving from the condition that the first (functional) derivative of an action, a quantity in the exponent of path integration, should vanish. Second we should check a positivity of the second derivative of the action at that point (or trajectory), that is, convexity of the action. If a solution meets this condition, designated as the stability condition\cite{rf:KOS2}, it is called the saddle point. Third, expanding the action around the saddle point and regarding terms greater than the third order derivative as perturbation we have a series expansion expressed by a loop expansion parameter, such as $\hbar$ or $1/N$. Almost every cases follow these standard procedures but sometimes there happens a situation that the second derivative itself also vanishes at the point, which is so called a caustics. The cure for this case has already been given, for example, in the textbook by Schulman\cite{rf:Schulman}: we should start with a point given on the condition that the second derivative vanishes. We then utilize the Airy integration, instead of the Gaussian integration in the standard case. A thorough study for caustics (quantum caustics) is seen in the recent works by Horie et al.\cite{rf:HMT}.

Another powerful technique exclusively used in the path integral formalism is the auxiliary field method(AFM); characterized by an additional degree inserted into the original expression in terms of a Gaussian integral to kill a four-body interaction. Although the method has a long history\cite{rf:GNKK, rf:FRAD}, it has recently been clarified that we can improve results when going into higher orders of the loop expansion in terms of the auxiliary field. It is important to note that even if the loop expansion parameter is not tiny but unity, results are still persistent\cite{rf:kashiwa}. There 
are many approximation schemes even under the path integral formalism, such as variational method\cite{rf:variational}, optimized perturbation theory\cite{rf:opv}, dilute gas approximation\cite{rf:SC}, valley method\cite{rf:vm}, etc., but "Simpler is Better".

In this paper, we study a 4-fermi interaction in 0-dimension as a prototype of the Nambu--Jona-Lasinio(NJL) model\cite{rf:NJL}. Although the NJL model was originally handled by the mean field technique, it is extensively analyzed with the aid of AFM\cite{rf:GNKK}. Our motivation is to establish the validity of AFM in NJL as was in the bosonic case\cite{rf:kashiwa}, however, we have found that the model itself posseses caustics. Therefore it is tempting to analyze it as a toy model to check the powerfulness of AFM even in 
a caustic example. By employing the standard prescription for the caustics, it is understood that AFM can give us a more accurate result if higher orders are taken into account.

The paper is organized as follows: in \S2, we propose our model and 
calculate its integration. We then introduce an auxiliary field and apply the saddle point method to that. In \S3 and \S4, we study cases of the 4-fermi coupling, $\lambda^2$, being posive and negative, respectively. The negative case, in \S4, contains caustics then we follow the standard procedure, utilizing the Airy integration instead of the Gaussian one. The final section is devoted to discussions.

\section{Grassmann Integral and Auxiliary Field Method}

The model considered here is
\begin{eqnarray}
  Z &=& \int d^N \mbit{\xi} d^N \mbit{\xi}^*  \exp \Bigl[ 
{-\omega(\mbit{\xi}^* \cdot \mbit{\xi})  +\frac{\lambda^2}{2N}}(\mbit{\xi}^*  \cdot 
\mbit{\xi})^2 \Bigl] \ , 
\label{gras}
\end{eqnarray}
where 
\begin{eqnarray}
  d^N \mbit{\xi} \equiv  d \xi_1 \cdot \cdot \cdot d \xi_N \ , \quad
  d^N \mbit{\xi}^* \equiv  d \xi^*_N \cdot \cdot \cdot d \xi^*_1 \ , \quad  
  (\mbit{\xi}^*  \cdot \mbit{\xi}) \equiv  \sum_{i=1}^N \xi^*_i  \xi_i \ ,
\end{eqnarray} 
and the coupling constant $\lambda^2$ is supposed real. We have introduced 
$2N$-Grassmann variables and our notaion is followed from the textbook of 
ref.\cite{rf:KOS}.

$Z$ in eq.(\ref{gras}) is exactly calculable by expanding the exponent and 
using the binomial theorem:
\begin{eqnarray}
  Z &=& \int d^N \mbit{\xi} d^N \mbit{\xi}^*
  \sum_{n=0}^{\infty} \frac{1}{n!}
  \bigl{\{}-\omega(\mbit{\xi}^* \cdot \mbit{\xi})
  +\frac{\lambda^2}{2N}(\mbit{\xi}^* \cdot \mbit{\xi})^2 \bigl{\}}^n 
  \nonumber \\
  &=& \int d^N \mbit{\xi} d^N \mbit{\xi}^*
  \sum_{n=0}^{\infty} 
  \sum_{r=0}^{n} \frac{1}{(n-r)!r!}
  (-\omega)^{n-r} (\frac{\lambda^2}{2N})^r (\mbit{\xi}^*  \cdot 
\mbit{\xi})^{n+r} \ .
  \nonumber 
\end{eqnarray}
Recalling the formula,
\begin{eqnarray}
\int d^N \mbit{\xi} d^N \mbit{\xi}^* (\mbit{\xi}^*  \cdot \mbit{\xi})^{m} 
=  (-1)^N N! \delta_{m N} \ ,
\end{eqnarray}
we obtain
\begin{eqnarray}
 Z =  \sum_{r=0}^{[\frac{N}{2}]}\frac{N!}{(N-2r)!r!} 
  \omega^{N-2r} (\frac{\lambda^{2}}{2N})^r \  ,  \label{exact} 
\end{eqnarray}
with $[\alpha]$ being the Gauss' symbol. In the following 
we compare this result with 
that of AFM with the help of the loop expansion, whose expansion parameter 
is now  $1/N$.

Introduce an auxiliary field, $y$,  to cancel the 
$(\mbit{\xi}^* \cdot \mbit{\xi} )^2$ term in eq.(\ref{gras}): 
insert the identity
\begin{eqnarray}
 1 = \int_{-\infty}^{\infty} \frac{d y}{\sqrt{2\pi}}  
    \exp \bigg[-\frac{1}{2} \big{\{} y + \frac{\lambda}{\sqrt{N}}
    (\mbit{\xi}^* \cdot \mbit{\xi} ) \big{\}}^2 \bigg] ,
\end{eqnarray}
to give 
\begin{eqnarray}  
  Z &=& \int_{-\infty}^{\infty} \frac{d y}
  {\sqrt{2\pi}}
  \int d^N \mbit{\xi} d^N \mbit{\xi}^*
  \exp \Bigl[-\frac{y^2}{2}
  -(\omega+\frac{\lambda}{\sqrt{N}} y)
  (\mbit{\xi}^* \cdot \mbit{\xi})
  \Bigl]  \nonumber \\
  &=& \int_{-\infty}^{\infty} \frac{d y}{\sqrt{2\pi}}
(\omega+\frac{\lambda}{\sqrt{N}} y)^N
  \exp \Bigl[-\frac{y^2}{2}  \Bigl]  \nonumber \\
&\stackrel{y \to \sqrt{N} y}{=}& 
  \int_{-\infty}^{\infty} \sqrt{\frac{N}{2\pi}} d y
  \exp [-N S(y)] \  , \label{base0}
\end{eqnarray}
with
\begin{eqnarray}
S(y) &\equiv& \frac{y^2}{2}-\ln (\omega+\lambda y) \ .
\end{eqnarray}

When $N$ goes larger (although we will assume that $N$ is not so large in 
the following), it is expected that the integral is dominated by the saddle 
point $y_0$ obeying both conditions such that 
\begin{eqnarray}
  0 = S^{(1)}(y_0) = y_0-\frac{\lambda}{\omega+\lambda y_0} \ ,
\label{classsol}
\end{eqnarray}
and
\begin{eqnarray}
S^{(2)}(y_0) > 0 \ ; \qquad \mbox{stability condition} \ . 
\label{stbCond}
\end{eqnarray}
Here and hereafter we employ the notation,
\begin{eqnarray}
S^{(n)}(y) &\equiv& \frac{d^n S(y)}{d y^n} \ . 
\end{eqnarray}

Write
\begin{eqnarray}
\Omega &\equiv& \omega+\lambda y_0 \ , \label{omega0}
\end{eqnarray}
to find that eq.(\ref{classsol}) becomes
\begin{eqnarray}
  \Omega-\omega-\frac{\lambda^2}{\Omega} = 0 \ , \label{classsol2}
\end{eqnarray} 
giving
\begin{eqnarray}
  \Omega = \frac{\omega \pm \sqrt{\omega^2+4\lambda^2}}{2} \ . 
\label{omega}
\end{eqnarray}

Expanding $S(y)$ around $y_0$, 
\begin{eqnarray}  
  Z = \!   \!  \!   \int_{-\infty}^{\infty} \!  \!  \!  
\sqrt{\frac{N}{2\pi}} d y
  \exp \bigg[  \!  -N \bigg{\{} S_0 + \frac{1}{2}S_0^{(2)}(y - y_0)^2  \!  
\!  + \sum_{n=3}^{\infty} \frac{1}{n!} S_0^{(n)}(y - y_0)^n \bigg{\}}  \!  
\bigg] \ , 
  \nonumber
\end{eqnarray}
then making a change of variable, $y-y_0 \mapsto y/\sqrt{N}$, 
we obtain
\begin{eqnarray}
  Z  = e^{-N S_0} 
  \int_{-\infty}^{\infty} \frac{d y}{\sqrt{2\pi}}
  \exp \Bigl[-\frac{S^{(2)}_0}{2}y^2 - \sum_{n=3}^{\infty}N^{1 - n/2} 
  \frac{S^{(n)}_0}{n!}y^n \Bigl]\ , \label{base}
\end{eqnarray}
with $ S^{(n)}_0 \equiv S^{(n)}(y_0)$. In view of eq.(\ref{base}) the terms 
of $n \geq 3$ becomes ineffective because of the factor $N^{1- n/2}$, which 
enable us to regard these as perturbation. (As stressed before, however, 
$N$ will be set in a small finite value in the following.) The first 6 terms 
are expressed as
\begin{eqnarray}
  S_0 &=& \frac{\lambda^2}{2 \Omega^2}-\ln \Omega , \label{s0} \\
  S_0^{(2)} &=& 1+\frac{\lambda^2}{(\omega+\lambda y_0)^2}
  = 1+\frac{\lambda^2}{\Omega^2}  \nonumber \\ 
  & = &  \frac{\sqrt{\omega^2 + 4 \lambda^2 }
    (\sqrt{\omega^2 + 4 \lambda^2 } \mp \omega) }{2 \lambda^2} \ , 
\label{s2} 
\end{eqnarray}
\begin{eqnarray}
  S_0^{(3)} &=& -\frac{2 \lambda^3}{(\omega+\lambda y_0)^3}
  = -\frac{2 \lambda^3}{\Omega^3} \ , \label{s3} \\
  S_0^{(4)} &=& \frac{6 \lambda^4}{(\omega+\lambda y_0)^4}
  = \frac{6 \lambda^4}{\Omega^4} \ , \label{s4} 
  \end{eqnarray}
\begin{eqnarray}
  S_0^{(5)} &=& -\frac{24 \lambda^5}{(\omega+\lambda y_0)^5}
  = -\frac{24 \lambda^5}{\Omega^5} \ , \label{s5} \\
  S_0^{(6)} &=& \frac{120 \lambda^6}{(\omega+\lambda y_0)^6}
  = \frac{120 \lambda^6}{\Omega^6} \ , \label{s6} 
\end{eqnarray}
respectively. The expansion up to the 3-loop correction is therefore read as
\begin{eqnarray}
  Z &=& e^{-N S_0} 
  \int_{-\infty}^{\infty} \frac{d y}{\sqrt{2\pi}}
  \exp \Bigl[ -\frac{S^{(2)}_0}{2}y^2 \Bigl]
  \Bigl[ 1 + \frac{1}{N} 
  \Bigl{\{} -\frac{S_0^{(4)}}{4!}y^4
  +\frac{(S_0^{(3)})^2}{2 (3!)^2} y^6 \Bigl{\}}  \nonumber \\
 &&  \hspace{-8mm}+ \frac{1}{N^2} \Bigl{\{} \! \!   -\frac{S_0^{(6)}}{6}y^6 
\!  \!  
  +  \!  \Bigl( \!  \frac{(S_0^{(4)})^2}{2 (4!)^2} \!  
  + \frac{S_0^{(3)} S_0^{(5)}}{3! 5!} \!  \Bigl) y^8 \!  \!  \! 
  -\frac{3 (S_0^{(3)})^2 S_0^{(4)}}{(3!)^3 4!}y^{10} \!  \! \! 
  +\frac{(S_0^{(3)})^4}{(3!)^4 4!}y^{12} \Bigl{\}}  \nonumber \\ 
  & + & \mathcal{O}(\frac{1}{N^3})\Bigl] \  . \label{base2}
\end{eqnarray}

\section{Calculation: $\lambda^2 > 0$ }

\begin{figure}[h]
 \begin{center}
\raisebox{0cm}{}
 \epsfxsize=10cm  \epsfbox{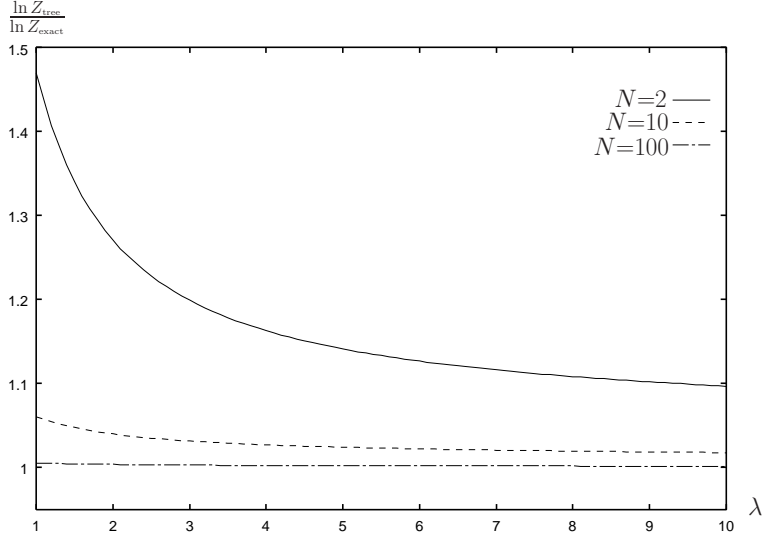}
 \end{center} 
\caption{\small The ratio; $\ln Z_{\rm tree}/\ln Z_{\rm exact}$ 
for $1< \lambda < 10$. 
The solid, the dotted and the dashed-dotted lines desiganate $N=2, 10$, and $100$ cases, respectively. When $N$ becomes larger the ratio closes to unity, showing reliablity of large $N$ approximation.}
\label{fig:1}
\end{figure}

In this case, eq.(\ref{omega}) reads
\begin{eqnarray}
  \Omega_{\pm} = \frac{1 \pm \sqrt{1+4\lambda^2}}{2}\ , \label{omega2}
\end{eqnarray}
so that the stability condition, eq.(\ref{stbCond}), with eq.(\ref{s2}) 
gives
\begin{eqnarray}
  S_0^{(2)} = 1+\frac{\lambda^2}{\Omega_{\pm}^2}
  = \frac{\sqrt{1 + 4 \lambda^2 }
    (\sqrt{1 + 4 \lambda^2 } \mp 1) }{2 \lambda^2} > 0\ ,
    \label{s22}
\end{eqnarray}
that is, both solutions meet the stability condition thus should be taken 
as the saddle points. (We have put $\omega = 1$ here and hereafter.) 
Therefore from eq.(\ref{base2}) the L-loop approximation for $Z$ is given by, 
\begin{eqnarray}
Z_{\mbox{L-loop}} = Z_{\mbox{L-loop}}^+ + Z_{\mbox{L-loop}}^- \ , 
\label{Lloop}
\end{eqnarray}
where
\begin{eqnarray}
Z_{\mbox{tree}}^{\pm } (\equiv Z_{\mbox{0-loop}}^{\pm }) \equiv  
\Omega_{\pm}^N \exp \Bigl[-\frac{N \lambda^2}{2 \Omega_{\pm}^2} \Bigl] \ ;  
\end{eqnarray}
and 
\begin{eqnarray}
  Z_{\mbox{3-loop}}^{\pm } &=&   \Omega_{\pm}^N 
  \exp  \Bigl[-\frac{N \lambda^2}{2 \Omega_{\pm}^2} \Bigl]
  \sqrt{\frac{\Omega_{\pm}^2}{\Omega_{\pm}^2  \! +  \! \lambda^2}}
  \nonumber \\
  & & \times  \Bigl[ 1 +\frac{1}{N} \Bigl{\{} -\frac{3 \lambda^4}
  {4(\Omega_{\pm}^2+\lambda^2)^2}
  +\frac{5 \lambda^6}{6(\Omega_{\pm}^2+\lambda^2)^3} \Bigl{\}}  \nonumber 
\\
  & & \hspace{-18mm} +{\frac{1}{N^2} \! \Bigl{ \{} -\frac{5 \lambda^6}
    {2(\Omega_{\pm}^2  \! +  \! \lambda^2)^3}
    + \frac{329 \lambda^8}{32(\Omega_{\pm}^2  \! +  \!\lambda^2)^4}} 
  -\frac{105 \lambda^{10}}{8(\Omega_{\pm}^2  \! +  \! \lambda^2)^5}
  +\frac{385 \lambda^{12}}
  {72(\Omega_{\pm}^2  \! +  \! \lambda^2)^6} \Bigl{\}} \Bigl] \ . 
  \label{conclusion0}
\end{eqnarray}
with the aid of eqs.(\ref{s0})$\sim$(\ref{s6}). In eq.(\ref{conclusion0}) 
we have the $1$- or the $2$-loop approximation if we neglect $\mathcal{O}(1/N)$ or 
$\mathcal{O}(1/N^2)$ term respectively.  

To see how the results depend on $N$, we plot the ratio, $\ln Z_{\rm tree}$ to $\ln Z_{\rm exact}$ in Fig.\ref{fig:1} for the cases of $N=2, 10$ and $100$ and $1< \lambda < 10$. (We should need the logarithm since Z itself becomes huge in $N=100$.)

\begin{table}

\begin{center}

\begin{tabular}{| c | c | c | c | c | c | }  \hline 
$\displaystyle{ \lambda }$  &  exact  & 
$\begin{array}{c}
\mbox{tree} \\
   \mbox{tree/ ex.}  
\end{array}$ & $\begin{array}{c}
\mbox{1-loop} \\
  \mbox{1-loop/ ex.}  
\end{array}$  & 
$\begin{array}{c}
\mbox{2-loop} \\
  \mbox{2-loop / ex.}  
\end{array}$  &   
$\begin{array}{c}
\mbox{3-loop} \\
  \mbox{3-loop / ex.}  
\end{array}$  \\   \hline 
$10^{-3}$ & 1.0000 & $\begin{array}{c}
  1.0000 \\   1.00  
\end{array}$ & $\begin{array}{c}
  1.0000 \\   1.00 
\end{array}$  & $\begin{array}{c}
  1.0000 \\   1.00                      
\end{array}$  &  $\begin{array}{c}
  1.0000 \\   1.00                      
\end{array}$   \\  \hline 
$10^{-2}$  & 1.0001 & $\begin{array}{c}
  1.0001  \\   1.00  
\end{array}$  & $\begin{array}{c}
  1.0001  \\   1.00 
\end{array}$  &  $\begin{array}{c}
  1.0001  \\   1.00  
\end{array}$   &  $\begin{array}{c}
 1.0001  \\   1.00  
\end{array}$  \\   \hline 
$10^{-1}$  & 1.0050 & $\begin{array}{c}
  1.0100  \\   1.00  
\end{array}$   &  $\begin{array}{c}
  1.0050  \\   1.00  
\end{array}$ & $\begin{array}{c}
  1.0050  \\   1.00  
\end{array}$ & $\begin{array}{c}
  1.0050  \\   1.00  
\end{array}$   \\  \hline  
$1 $  & 1.5000 & $\begin{array}{c}
  1.8147  \\   1.21  
\end{array}$ &  $\begin{array}{c}
  1.5346 \\   1.02  
\end{array}$  &  $\begin{array}{c}
  1.5039 \\   1.00  
\end{array}$  &   $\begin{array}{c}
  1.4996  \\   1.00  
\end{array}$ \\  \hline
$10 $  & 51.000  & $\begin{array}{c}
  74.679    \\   1.46  
\end{array}$ &  $\begin{array}{c}
  53.050  \\   1.04  
\end{array}$  &  $\begin{array}{c}
  50.866 \\  0.997  
\end{array}$   &  $\begin{array}{c}
  50.902 \\  0.998
\end{array}$   \\
\hline
 \end{tabular}

\end{center}
\caption{\small The exact value as well as that of $Z_{\rm L-loop}$  and the ratio, $Z_{\rm L-loop}/ Z_{\rm exact}$, are listed in $N=2$ case for $10^{-3} \leq \lambda \leq 10$. The 2-loop results almost reproduce the exact value.} 
\label{tab:1}
\end{table}

\begin{figure}[h]
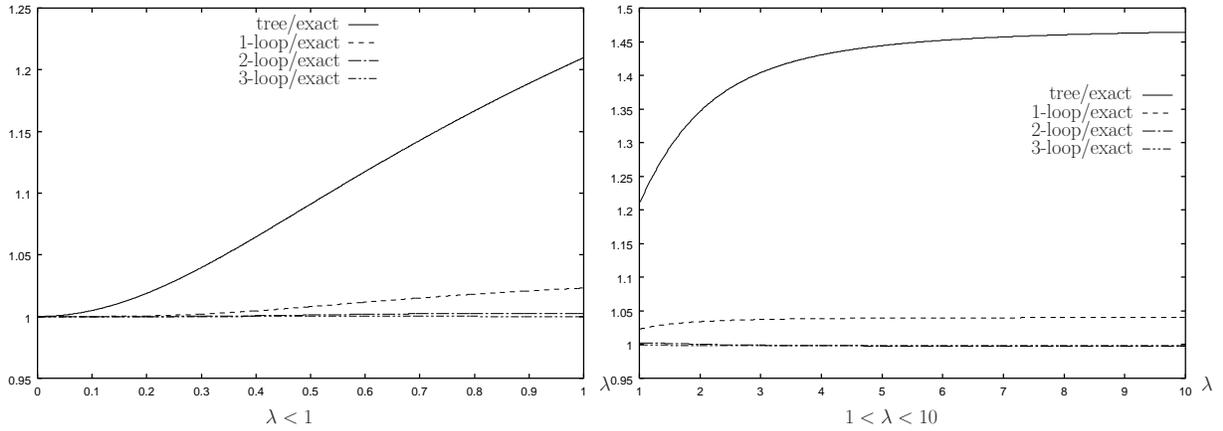

\begin{center}
\raisebox{0cm}{} 
\epsfxsize=8cm  \epsfbox{g+small.epsi}  \epsfxsize=8cm \epsfbox{g+large.epsi}  
\end{center}
 \caption{\small The ratio; $Z_{\rm L-loop}/ Z_{\rm exact}$ in $N=2$ case for $10^{-3} \leq \lambda \leq 10$. The solid, dotted, dashed-dotted and dashed-double-dotted lines correspond to the tree, 1-, 2-, and 3-loop results, respectively.}
\label{fig:2}
\end{figure}

\begin{figure}[h]
\begin{center}
\raisebox{0cm}{} 
\epsfxsize=8cm  \epsfbox{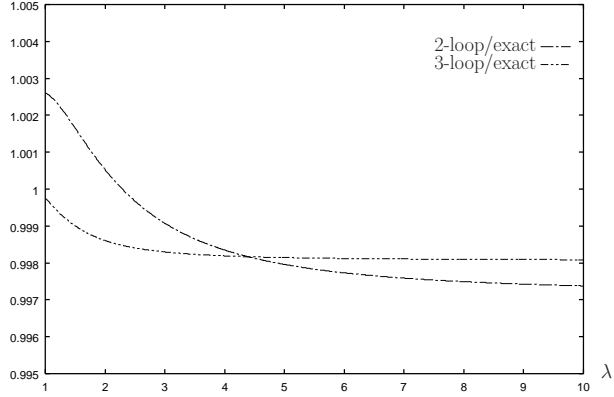}
\end{center}
\caption{\small The 2- and 3-loop results to the exact value are plotted. The 2-loop dominates the 3-loop between $\lambda \sim 2$ and $\lambda \sim 4$, implying that the loop expansion is merely an aymptotic expansion.}
\label{fig:3}
\end{figure}

From Fig.\ref{fig:1}, it is obvious that the tree result becomes exact under $N \mapsto \infty $, which guarantees the large $N$ expansion. However, as is the bosonic case\cite{rf:kashiwa}, even when $N$ is small, the smallest, $N=2$, in this case, we can get a better answer by taking higher loops into account: the results are listed in Table \ref{tab:1} where the ratio $Z_{\rm L-loop}/ Z_{\rm exact}$ is listed for $L=0,1,2,3$. In spite of the fact that at $\lambda=10$ the error is reaching $\sim 50\%$ in the tree approximation, it is cured to $4 \%$ in the 1-loop correction and almost to the exact value in the 2-loop. The same ratio is plotted in Figs.\ref{fig:2}, which shows that higher the loop we go the better the results are. As was discussed in the previous work\cite{rf:kashiwa}, however, since the loop expansion is merely an asymptotic expansion, it is not always true that a higher loop effect can be more accurate; whose fact can be seen in Fig.\ref{fig:3}. The 2-loop result becomes better than that of the 3-loop in $  2 \  \raisebox{-1mm}{$\stackrel{ <}{ \sim}$} \  \lambda \ \raisebox{-1mm}{$\stackrel{<}{\sim}$} \ 4$.

\section{Calculation: $\lambda^2 < 0$ ; caustic case}

Next, we consider the $\lambda^2 < 0$ case. Write $\lambda^2 = 
-|\lambda|^2$ to find $\Omega$ and $S_0^{(2)}$ are 
\begin{eqnarray}
  \Omega_{\pm} &=& \frac{1 \pm \sqrt{1-4|\lambda|^2}}{2} \ , \label{omega3} 
\\
  S_0^{(2)} &=& 1-\frac{|\lambda|^2}{\Omega_{\pm}^2}
  = \frac{\sqrt{1 - 4 |\lambda|^2 }
    (\pm 1 - \sqrt{1 - 4 |\lambda|^2 }) }{2 |\lambda|^2} \ ,  \label{s23}
\end{eqnarray}
respectively. From eq.(\ref{s23}), we find $S_0^{(2)} = 0$ when $|\lambda| = 1/2$, that is, caustics. We cannot proceed anymore around this value with the standard recipe in terms of Gaussian integration.  We then divide the region of $|\lambda|$ into three segments: 
i) $|\lambda| < 1/2$ , ii) $|\lambda| > 1/2$, and iii) $|\lambda| \approx 1/2$, and designate i) and ii) as the Gaussian region and iii) as the Airy region. The reason is that in i) and ii) we can proceed as usual but in iii) we should adopt the Airy integral, as will be shown in the following.

\subsection{Gaussian region I: $|\lambda| < 1/2$}

Let us start with the case, $|\lambda| < 1/2$. In view of eq.(\ref{s23}), only $\Omega_+$ is the saddle point; since $\Omega_-$ does not satisfy the stability condition. A general expression, eq.(\ref{base2}), then reads
\begin{eqnarray}
  Z &=& \Omega_+^N \exp 
  \Bigl[\frac{N |\lambda|^2}{2 \Omega_+^2} \Bigl]
  \sqrt{\frac{\Omega_+^2}{\Omega_+^2-|\lambda|^2}}
  \Bigl[ 1  \nonumber \\
  & & +  \frac{1}{N} \Bigl{\{} -\frac{3 |\lambda|^4}
  {4(\Omega_+^2-|\lambda|^2)^2}
  -\frac{5 |\lambda|^6}{6(\Omega_+^2-|\lambda|^2)^3} \Bigl{\}}  
  \nonumber \\
  & & +{\frac{1}{N^2} \Bigl{\{} \frac{5 |\lambda|^6}
    {2(\Omega_+^2-|\lambda|^2)^3}
    +\frac{329 |\lambda|^8}{32(\Omega_+^2-|\lambda|^2)^4}}   \nonumber \\
  & & \hspace{7mm} +\frac{105 |\lambda|^{10}}{8(\Omega_+^2-|\lambda|^2)^5}
  +\frac{385 |\lambda|^{12}}
  {72(\Omega_+^2-|\lambda|^2)^6} \Bigl{\}}   \nonumber \\
  & & \hspace{35mm}+ \mathcal{O}(\frac{1}{N^3}) \Bigl] \ . 
\label{conclusion1}
\end{eqnarray}

\subsection{Gaussian region II: $|\lambda| > 1/2$}

In this case, eqs.(\ref{omega3}), (\ref{s23}) are rewritten as 
\begin{eqnarray}
  \Omega_{\pm} &=& \frac{1 \pm i \sqrt{4|\lambda|^2-1}}{2} \ , 
\label{omega4}  \\
  S_0^{(2)} &=&  \frac{ i \sqrt{4 |\lambda|^2 -1}
    (\pm 1 - i \sqrt{4 |\lambda|^2 -1}) }{2 |\lambda|^2} \nonumber \\
 &  = & \frac{\sqrt{4 |\lambda|^2 -1}}{|\lambda|}
  \bigg(\frac{\sqrt{4 |\lambda|^2 -1}}{2 |\lambda|} 
  \pm i \frac{1}{2 |\lambda|} \bigg) \ .  \label{s24}
\end{eqnarray}
As $S_0^{(2)}$ is complex, the integration range should be analytically 
continued from the real axis to the contour satisfied with conditions,
\begin{eqnarray}
  \mbox{Re}[S_0^{(2)} y^2] > 0 \  , \qquad  
  \mbox{Im}[S_0^{(2)} y^2] = 0 \ .  \label{imag}
\end{eqnarray}
To this end, write
\begin{eqnarray}
  S_0^{(2)} &=& R e^{\pm i \theta} \ ,  \label{s25} 
\end{eqnarray}
with
\begin{eqnarray}
  R \equiv \frac{\sqrt{4|\lambda|^2 - 1}}{|\lambda|} \ ,  \quad 
  \tan \theta \equiv  \frac{1}{\sqrt{4|\lambda|^2 - 1}}  \ ; \quad   (0 < 
\theta < \frac{\pi}{2}) \ ,   \label{tan}
\end{eqnarray}
and 
\begin{eqnarray}
  y &=& r e^{ i \varphi},  \label{y1} 
\end{eqnarray}
to find 
\begin{eqnarray}
  S_0^{(2)} y^2 &=& r^2 R e^{2 i \varphi \pm i \theta}.  
\label{s25y}\end{eqnarray}
The one of the conditions (\ref{imag}) thus reads
\begin{eqnarray}
 0= \mbox{Im}[S_0^{(2)} y^2] = r^2 R \sin (2 \varphi \pm \theta ) 
  \ ,  \label{imag2}
\end{eqnarray}
namely,
\begin{eqnarray}
  \varphi &=& \mp \frac{\theta}{2} \ ,  \label{varphi} 
\end{eqnarray}
and the rest of it gives
\begin{eqnarray}
\mbox{Re}[S_0^{(2)} y^2] = r^2 R \cos (2 \varphi \pm \theta ) 
  = r^2 R  \  ,  \label{R} 
\end{eqnarray}
which is apparently positive. Therefore the stability condition is 
established for both cases $\Omega_{+}$ and $\Omega_{-}$. The quantities, 
eqs.(\ref{omega4}),(\ref{s0}), and (\ref{s3}) $\sim$ (\ref{s6}), 
read
\begin{eqnarray}
  \Omega_{\pm} &=& \pm i |\lambda| ( \frac{\sqrt{4|\lambda|^2}-1}{2 
|\lambda|}
  \mp \frac{1}{2 |\lambda|} ) = |\lambda| e^{\pm i (\frac{\pi}{2} - 
\theta)} \  ,
  \label{omega5} \\
  S_0 &=& -\frac{|\lambda|^2}{2 |\lambda|^2 e^{\pm i (\pi - 2 \theta)}}
  -\ln (|\lambda| e^{\pm i (\frac{\pi}{2} - \theta)}) \nonumber \\
   &=& \frac{1}{2} \cos 2 \theta -\ln |\lambda| 
  \mp i (\frac{\pi}{2} - \theta - \frac{1}{2} \sin 2 \theta) \ , 
\label{s02} 
\\
  S_0^{(3)} &=& \frac{2 i |\lambda|^3}{|\lambda|^3 
    e^{\pm i (\frac{3}{2} \pi - 3 \theta)}}
  = \mp 2 e^{\pm 3 i \theta} \  , \label{s32} \\
  S_0^{(4)} &=& \frac{6 |\lambda|^4}{|\lambda|^4 
    e^{\pm i (2 \pi - 4 \theta)}}
  = 6 e^{\pm 4 i \theta} \ , \label{s42} \\
  S_0^{(5)} &=& - \frac{24 i |\lambda|^5}{|\lambda|^5 
    e^{\pm i (\frac{5}{2} \pi - 5 \theta)}}
  = \mp 24 e^{\pm 5 i \theta} \ , \label{s52} \\
  S_0^{(6)} &=& - \frac{120 |\lambda|^6}{|\lambda|^6 
    e^{\pm i (3 \pi - 6 \theta)}}
  = 120 e^{\pm 6 i \theta}  \ . \label{s62} 
\end{eqnarray} 
As was in eq.(\ref{Lloop}) both solutions $\Omega_{\pm}$ come into play but they are complex conjugate each other, so that eq.(\ref{Lloop}) reads
\begin{eqnarray}
  Z_{\mbox{L-loop}} = Z_{\mbox{L-loop}}^+ + Z_{\mbox{L-loop}}^- 
  = 2 \mbox{Re} \left[ Z_{\mbox{L-loop}}^+ \right] \ . \label{conc3} 
\end{eqnarray}
Therefore
\begin{eqnarray}
  Z_{\mbox{tree}} \equiv  2 |\lambda|^N e^{-\frac{N}{2} \cos 2 \theta} 
  \cos N \bigg( \frac{\pi}{2} - \theta - \frac{1}{2} \sin 2 \theta \bigg) \ 
,
  \label{conc3tree} 
\end{eqnarray}
and
\begin{eqnarray}
  Z_{\mbox{3-loop}} &\equiv & 
  \frac{2 |\lambda|^N}{\sqrt{R}} e^{-\frac{N}{2} \cos 2 \theta} 
  \bigg[ \cos \bigg{\{}N \bigg( \frac{\pi}{2} - \theta - \frac{1}{2} 
  \sin 2 \theta \bigg)-\frac{\theta}{2} \bigg{\}} \nonumber \\
  &+& \frac{1}{N} \bigg{\{} -\frac{3}{4 R^2} 
  \cos \bigg{\{}N \bigg( \frac{\pi}{2} - \theta - \frac{1}{2} 
  \sin 2 \theta \bigg) + \frac{3}{2} \theta \bigg{\}} \nonumber \\
  &&+ \frac{5}{6 R^3} 
  \cos \bigg{\{}N \bigg( \frac{\pi}{2} - \theta - \frac{1}{2} 
  \sin 2 \theta \bigg)+ \frac{5}{2} \theta \bigg{\}}
  \bigg{\}} \nonumber \\
  &+& \frac{1}{N^2} \bigg{\{} -\frac{5}{2 R^3} 
  \cos \bigg{\{}N \bigg( \frac{\pi}{2} - \theta - \frac{1}{2} 
  \sin 2 \theta \bigg)+\frac{5}{2} \theta \bigg{\}} \nonumber \\
  &&+ \frac{329}{32 R^4} 
  \cos \bigg{\{}N \bigg( \frac{\pi}{2} - \theta - \frac{1}{2} 
  \sin 2 \theta \bigg)+\frac{7}{2} \theta \bigg{\}} \nonumber \\
  &&- \frac{105}{8 R^5} 
  \cos \bigg{\{}N \bigg( \frac{\pi}{2} - \theta - \frac{1}{2} 
  \sin 2 \theta \bigg)+\frac{9}{2} \theta \bigg{\}} \nonumber \\
  &&+ \frac{385}{72 R^6} 
  \cos \bigg{\{}N \bigg( \frac{\pi}{2} - \theta - \frac{1}{2} 
  \sin 2 \theta \bigg)+\frac{11}{2} \theta \bigg{\}}
  \bigg{\}} \bigg]
 \label{conc33}  \ , 
\end{eqnarray}
where use has been made of the Gaussian integration formula, 
\begin{eqnarray}
  \int_{-\infty}^{\infty} \frac{d y}{\sqrt{2\pi}} y^{2n}
  \exp \Bigl[ -\frac{S^{(2)}_0}{2}y^2 \Bigl]
&=& e^{\mp i \frac{2n+1}{2} \theta} 
\int_{-\infty}^{\infty} \frac{d r}{\sqrt{2\pi}} r^{2n}
  \exp \Bigl[ -\frac{1}{2} R r^2 \Bigl] \nonumber \\
&=& \frac{(2n-1)!!}{R^{n+\frac{1}{2}}}
e^{\mp i \frac{2n+1}{2} \theta} \quad  (n \geq 0) \  . \label{gauss}
\end{eqnarray}
Here in eq.(\ref{conc33}), $ Z_{\mbox{1-loop}}$ and $ Z_{\mbox{2-loop}}$ approximations are obtained by dropping terms of $\mathcal{O}(1/N)$ and $\mathcal{O}(1/N^2)$ respectively.

\subsection{Airy region: $|\lambda| \approx 1/2$}

As was mentioned several times before, the Gaussian integtration cannot be
used in this region because $S_0^{(2)} \approx 0$; instead, we should 
employ a new approximation based on the Airy integral: first find the solution 
$\tilde{y}$ of 
\begin{eqnarray}
S_0^{(2)} = 1-\frac{|\lambda|^2}{(\omega+i |\lambda| \tilde{y})^2} = 0 \ , 
\label{s2cau}
\end{eqnarray}
instead of $S^{(1)}(y) = 0$ \ . 
As in eq.(\ref{omega0}), define
\begin{eqnarray}
\tilde{\Omega} &\equiv& \omega+i |\lambda| \tilde{y} \ , \label{omega0cau}
\end{eqnarray} 
to obtain
\begin{eqnarray}
\tilde{\Omega}_{\pm} = \pm |\lambda| \ . \label{omegacau}
\end{eqnarray} 
Write 
\begin{eqnarray}
  \tilde{S}^{(n)} \equiv S^{(n)} (\tilde{y}) \  ,
\end{eqnarray}
(again, putting $\omega = 1$) then, 
\begin{eqnarray}
  \tilde{S}^{(1)}_{\pm} = \frac{i}{|\lambda|}(1 \mp |\lambda|) 
  - \frac{i |\lambda|}{(\pm |\lambda|)} 
  = \frac{i}{|\lambda|}(1 \mp 2|\lambda|) \ . \label{s1cau}
\end{eqnarray}
We should adopt $\tilde{\Omega}_+$ as a suitable solution; because which 
connects smoothly with the saddle point: $\tilde{\Omega}_+$ yields 
$\tilde{S}^{(1)}_+ = 0$. At this point
\begin{eqnarray}
  \tilde{S} = \frac{1}{2} \bigg{\{} \frac{i}{|\lambda|}(1-|\lambda|) 
\bigg{\}}^2 \! \!  \! \! \! \! \!  &  -    &  \! \! \! \! \ln |\lambda| 
  = -\frac{(1 - |\lambda|)^2}{2|\lambda|^2} - \ln |\lambda| \ , 
\label{s0cau} 
\\
 \tilde{S}^{(3)} = \frac{2 i |\lambda|^3}{|\lambda|^3} = 
2i \ ,  & & \quad   \tilde{S}^{(4)} = \frac{6 |\lambda|^4}{|\lambda|^4} = 6 
\ ,\label{s4cau} 
\\
  \tilde{S}^{(5)} \! \! \! \! \!  &=& \! \! \! -\frac{24 i |\lambda|^5}{|\lambda|^5} = -24i \ 
.\label{s5cau}
\end{eqnarray}
Second, expand the exponent $S(y)$ around $\tilde{y}$ and make a change of 
variable,
\begin{eqnarray}
  y-\tilde{y} = i \bigl( \frac{2}{N \tilde{S}^{(3)}} \bigl)^{\frac{1}{3}}t
  =-\frac{t}{N^{\frac{1}{3}}} \ , \label{vari} 
\end{eqnarray} 
to find 
\begin{eqnarray}
  Z &=& e^{-N \tilde{S}} 
    \int_{-\infty}^{\infty} \frac{N^{\frac{1}{6}}}{\sqrt{2\pi}} d t        
      \exp \Bigl[ \frac{i t^3}{3}+i a t - \sum_{n=4}^{\infty} 
  \frac{ (-1)^n \tilde{S}^{(n)}}{n! N^{\frac{n}{3}-1}} t^n\Bigl], 
\label{airy}       
\end{eqnarray}
where
\begin{eqnarray}
  a \equiv  \frac{N^{\frac{2}{3}}}{|\lambda|}(1 - 2|\lambda|) \ . \label{a}
\end{eqnarray}
Note $a \sim \mathcal{O}(1)$ even under $N \mapsto \infty$ since $1- 2|\lambda| 
\approx 0$ in this region.  Therefore, third we keep the first two terms in the exponent and regard others as perturbation to obtain
\begin{eqnarray}
  Z &=& |\lambda|^N \exp 
  \Bigl[N \frac{(1 - |\lambda|)^2}{2 |\lambda|^2} \Bigl]
  \sqrt{\frac{2}{\pi}}N^{\frac{1}{6}}  \nonumber \\
 & & \times  \Bigl[ \int_0^\infty dt \cos \Bigl[\frac{t^3}{3}+at \Bigl] 
\nonumber \\
  & & -\frac{1}{4N^{\frac{1}{3}}} \int_0^\infty dt \,t^4 \cos 
\Bigl[\frac{t^3}{3}+at \Bigl]   \nonumber \\
  &&  + \frac{1}{N^{\frac{2}{3}}} 
  \Bigl{\{} \frac{1}{32}
  \int_0^\infty dt \,t^8 \cos \Bigl[\frac{t^3}{3}+at \Bigl]
  +\frac{1}{5}
  \int_0^\infty dt \,t^5 \sin \Bigl[\frac{t^3}{3}+at \Bigl] \Bigl{\}} 
  \nonumber \\
 & &  \hspace{18mm}+\mathcal{O}(\frac{1}{N})\Bigl] \ . \label{coacau}
\end{eqnarray}

\begin{figure}[h]
\begin{center}
\raisebox{0cm}{} 
\epsfxsize=8cm \epsfbox{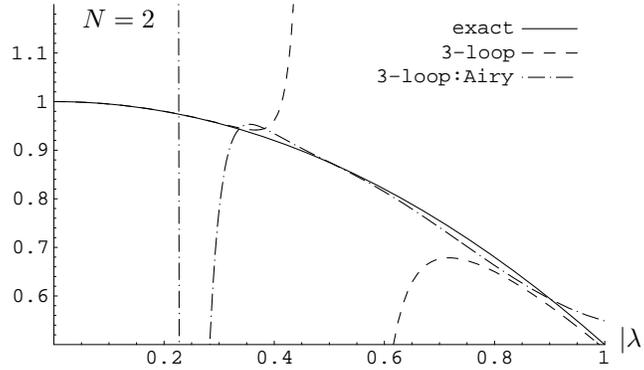} 
\put(3,9){$|\lambda|$}
\put(-200,130){$N=2$}
\caption{\small $N=2$ case for $10^{-3} \leq |\lambda| \leq 1 $: 
the solid, dotted, and dashed-dotted lines correspond to the exact, 
the Gaussian 3-loop, the (Airy) 3-loop results, respectively.  
The Airy curve drops down at $|\lambda| \sim 3.5$, 
reaches the bottom $\sim -1.88$ at $|\lambda| =0.238$ 
and goes up to infinity as $|\lambda| \mapsto 0$.}
\label{fig:4} 
\end{center}
\end{figure}

\begin{figure}[h]
\begin{center}
\raisebox{0cm}{} 
\epsfxsize=8cm \epsfbox{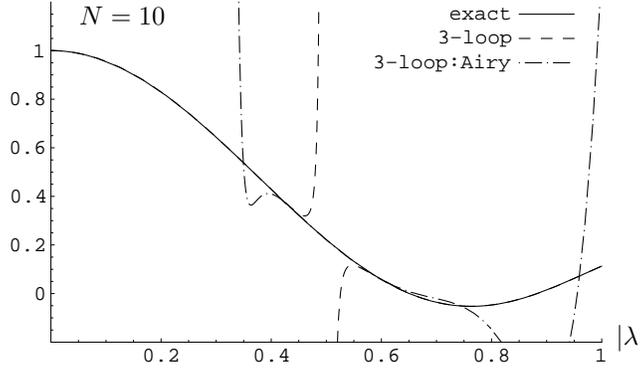} 
\put(3,9){$|\lambda|$}
\put(-200,130){$N=10$}
\end{center}
\caption{\small The same plot for $N=10$. The 3-loop Airy at $ 0.4 < |\lambda| < 0.6 $ as well as the 3-loop Gaussian at $|\lambda| < 0.4$ and $|\lambda| > 0.6$ fits excellently everywhere, showing reliablity of large $N$ approximation.}
\label{fig:5}    
\end{figure}

This expansion is not a loop expansion given in terms of $1/N$ but of 
$1/N^{1/3}$, however, we adopt the same terminology of the loop-expansion as the Gaussian case, that is, designate the first factor as a tree, the second integral of $\mathcal{O}(1)$ as a 1-loop, and those of $\mathcal{O}(1/N^{1/3})$ and $\mathcal{O}(1/N^{2/3})$ as 2- and 3-loop, respectively. We thus write 
$Z_{\mbox{3-loop}}$ as an approximation keeping terms up to $\mathcal{O}(1/N^{2/3})$ in 
eq.(\ref{coacau}). Finally, we utilize the Airy integral\cite{rf:Airy}
\begin{eqnarray}
 {\rm Ai}(a) = \frac{1}{\pi } \int_{0}^{\infty} d t        
  \cos \Bigl[ \frac{ t^3}{3}+ a t \Bigl] \ ; \qquad  {\rm Im}[a] = 0 \ ,
  \label{Airy}
\end{eqnarray} 
with ${\rm Ai}(a)$ being the Airy function, instead of the Gaussian 
integral. Therefore
\begin{eqnarray}
Z_{\mbox{3-loop}} &=& |\lambda|^N \exp 
  \Bigl[N \frac{(1 - |\lambda|)^2}{2 |\lambda|^2} \Bigl]
  \sqrt{2 \pi}N^{\frac{1}{6}}  \nonumber \\
 &\times &   \Bigl[ {\rm Ai}(a)  -\frac{1}{4N^{\frac{1}{3}}}{\rm 
Ai}^{(4)}(a)   + \frac{1}{N^{\frac{2}{3}}} 
  \Bigl{\{} \frac{1}{32}
  {\rm Ai}^{(8)}(a)
  - \frac{1}{5}
 {\rm Ai}^{(5)}(a) \Bigl{\}} \Bigl] \ , 
 \label{airy3loop}
\end{eqnarray}
with
\begin{eqnarray}
{\rm Ai}^{(n)}(a) \equiv \frac{d^n}{da^n}{\rm Ai}(a)  \ .
\end{eqnarray}
In eq.(\ref{airy3loop}), if we drop $\mathcal{O}(1)$ (${\rm Ai}(a)$) term, we have 
the tree result, and if drop $\mathcal{O}(1/N^{1/3})$ term, we have the 1-loop result 
and so on. By noting that 
\begin{eqnarray}
{\rm Ai}^{(2)}(a) = a {\rm Ai}(a) \ ,
\end{eqnarray}
which is immediately derived from the definition, eq.(\ref{Airy}), the expression, eq.(\ref{airy3loop}), can be given by the Airy function and its first derivative. After that we estimate it numerically using Mathematica. 

\begin{table}
\begin{center}
\begin{tabular}{| c | c | c | c | c | c | }  \hline 
$\displaystyle{ |\lambda| }$  &  exact  & 
$\begin{array}{c}
\mbox{tree} \\
   \mbox{tree/ ex.}  
\end{array}$ & $\begin{array}{c}
\mbox{1-loop} \\
  \mbox{1-loop/ ex.}  
\end{array}$  & 
$\begin{array}{c}
\mbox{2-loop} \\
  \mbox{2-loop / ex.}  
\end{array}$  &   
$\begin{array}{c}
\mbox{3-loop} \\
  \mbox{3-loop / ex.}  
\end{array}$  \\   \hline 
0.3 & 0.955 & $\begin{array}{c}
  20.832 \\   21.8  
\end{array}$ & $\begin{array}{c}
  1.7111 \\   1.79 
\end{array}$  & $\begin{array}{c}
  1.2467 \\   1.31                      
\end{array}$  &  $\begin{array}{c}
  0.7789 \\   0.816                      
\end{array}$   \\  \hline 
0.4  & 0.92 & $\begin{array}{c}
  1.5180  \\   1.65  
\end{array}$  & $\begin{array}{c}
  0.7305  \\   0.794 
\end{array}$  &  $\begin{array}{c}
  0.9566  \\   1.04  
\end{array}$   &  $\begin{array}{c}
 0.9341  \\   1.02  
\end{array}$  \\   \hline 
0.5  & 0.875 & $\begin{array}{c}
  0.6796  \\   0.777  
\end{array}$   &  $\begin{array}{c}
  0.6788  \\   0.776 
\end{array}$ & $\begin{array}{c}
  0.8752  \\   1.00  
\end{array}$ & $\begin{array}{c}
  0.8752  \\   1.00  
\end{array}$   \\  \hline  
0.6  & 0.82 & $\begin{array}{c}
  0.5615  \\   0.684  
\end{array}$ &  $\begin{array}{c}
  0.7608 \\   0.928  
\end{array}$  &  $\begin{array}{c}
  0.8419 \\   1.03  
\end{array}$  &   $\begin{array}{c}
  0.8145  \\   0.993  
\end{array}$ \\  \hline
0.7  & 0.755  & $\begin{array}{c}
  0.5888    \\   0.780  
\end{array}$ &  $\begin{array}{c}
  0.8820  \\   1.17  
\end{array}$  &  $\begin{array}{c}
  0.7758 \\  1.03  
\end{array}$   &  $\begin{array}{c}
  0.7411 \\  0.982
\end{array}$   \\ \hline
0.8  & 0.68  & $\begin{array}{c}
  0.6813    \\   1.00  
\end{array}$ &  $\begin{array}{c}
  1.0105  \\   1.49  
\end{array}$  &  $\begin{array}{c}
  0.6494 \\  0.955  
\end{array}$   &  $\begin{array}{c}
  0.6628 \\  0.975
\end{array}$   \\
\hline
 \end{tabular}
\end{center}

\caption{\small The Airy loop expansion in $N=2$ case 
for $0.3 \leq |\lambda| \leq 0.8$. As is the case of the Gaussian, 
the 3-loop correction improves the results but in $0.4 \leq |\lambda| \leq 0.7$ the 2-loop correction is sufficient.} 
\label{tab:3}
\end{table}

In order to show the $N$-dependence of the results, we first plot the whole region, $10^{-3} \leq |\lambda| \leq 1$, in Fig.\ref{fig:4} for $N=2$ and in Fig.\ref{fig:5} for $N=10$. As is seen from the graphs,  fitting of $N=10$ in Fig.\ref{fig:5} becomes better than that of $N=2$ in Fig.\ref{fig:4}, so that in the following we deal only with $N=2$ to check the reliablity of AFM. It is also apparent that the Gaussian loop expansion matches with the exact value for $|\lambda| > 1$ thus here and hereafter we concentrate only on the region, $10^{-3} \leq |\lambda| \leq 1$\footnote{See the dotted lines in Fig.\ref{fig:4}; it almost overlaps the solid line and in Fig.\ref{fig:5} it perfectly overlaps.}.

To see efficiency of the Airy loop expansion as in the Gaussian case, $\lambda^2 >0$ , of Table \ref{tab:1}, we list the Airy L-loop approximation for $0.3 \leq |\lambda| \leq 0.8$ in Table \ref{tab:3}. From this, although the 3-loop approximation does work well for $0.4 \leq |\lambda| \leq 0.8$, the Airy 2-loop correction is enough to produce an accurate value for $  0.4 \  \raisebox{-1mm}{$\stackrel{<}{\sim}$} \ |\lambda| \ \raisebox{-1mm}{$\stackrel{<}{\sim}$} \  0.7$. The situation is similar in $\lambda^2 >0$. Therefore the final result with the Airy 3-loop in $N=2$ is seen in Table \ref{tab:2}. From this, if we incoperate the Airy region by switching back and forth, from and to the Gaussian region, around $|\lambda| \sim 0.4$ and  $\sim 0.7$, we find that errors remain within $2\%$.

\begin{table}

\begin{center}

\begin{tabular}{| c | c | c | c || c | c | c | c | }  \hline 
    $\displaystyle{ |\lambda| }$  &  exact  & 
    $\begin{array}{c}
      \mbox{Gauss} \\
    \!  \!  \!  \! \mbox{Gauss/ ex.}  \!  \!  \!  \! 
    \end{array}$ & $\begin{array}{c}
      \mbox{Airy} \\
     \!  \!  \!  \!   \mbox{Airy/ ex.}   \!  \!  \!  \!
    \end{array}$  & 
    $\displaystyle{ |\lambda| }$  &  exact  & 
    $\begin{array}{c}
      \mbox{Gauss} \\
    \!  \!  \!  \! \mbox{Gauss/ ex.}  \!  \!   \!   \! 
 \end{array}$ & $\begin{array}{c}
   \mbox{Airy} \\
   \!  \!  \!  \!  \mbox{Airy/ ex.}   \!  \!  \!  \! 
 \end{array}$ \\   \hline 
 $10^{-3}$ & 1.0000 & $\begin{array}{c}
   1.0000 \\   1.00  
 \end{array}$ & $\begin{array}{c}
  \mbox{---} \\ \mbox{---}
 \end{array}$ & 
 0.5 & 0.8750 & $\begin{array}{c}
   \infty \\ \mbox{---}                        
\end{array}$  &  $\begin{array}{c}
  0.8752 \\   1.00                      
\end{array}$   \\  \hline 
 $10^{-2}$ & 1.0000 & $\begin{array}{c}
   1.0000 \\   1.00  
 \end{array}$ & $\begin{array}{c}
   \mbox{---} \\   \mbox{---} 
 \end{array}$ & 
 0.6 & 0.8200 & $\begin{array}{c}
   0.3577 \\   0.436                      
\end{array}$  &  $\begin{array}{c}
  0.8145 \\   0.993                      
\end{array}$   \\  \hline 
 $10^{-1}$  & 0.9950 & $\begin{array}{c}
  0.9950  \\   1.00  
\end{array}$  & $\begin{array}{c}
  \mbox{---}  \\   \mbox{---} 
\end{array}$  &
0.7 & 0.7550 & $\begin{array}{c}
  0.6764  \\   0.896  
\end{array}$   &  $\begin{array}{c}
 0.7411  \\   0.982  
\end{array}$  \\   \hline 
0.2 & 0.9800 & $\begin{array}{c}
  0.9800  \\   1.00  
\end{array}$   &  $\begin{array}{c}
  \mbox{---}  \\   \mbox{---}  
\end{array}$ & 
0.8 & 0.6800 & $\begin{array}{c}
  0.6500  \\   0.956  
\end{array}$ & $\begin{array}{c}
  0.6628  \\   0.975  
\end{array}$   \\  \hline  
0.3  & 0.9550 & $\begin{array}{c}
  0.9556  \\   1.00  
\end{array}$ &  $\begin{array}{c}
  0.7789 \\  0.816   
\end{array}$  &
0.9 & 0.5950 &   $\begin{array}{c}
  0.5793 \\   0.974  
\end{array}$  &   $\begin{array}{c}
   0.5941   \\  0.999    
\end{array}$ \\  \hline
0.4 & 0.9200  & $\begin{array}{c}
  0.9614    \\   1.04  
\end{array}$ &  $\begin{array}{c}
  0.9341  \\   1.02  
\end{array}$  & 
1.0 & 0.5000 &  $\begin{array}{c}
  0.4904 \\  0.981  
\end{array}$   &  $\begin{array}{c}
  0.5486 \\  1.10
\end{array}$   \\
\hline
\end{tabular}
 
\end{center}
\caption{\small Results in $N=2$: those obtained by the Gaussian 3-loop expansion are listed in $10^{-3} \leq |\lambda| \leq 1$ except $|\lambda| \neq 1/2$. Also those by the Airy 3-loop in $0.3 \leq |\lambda| \leq 1$. The lower quantities correspond to the ratio of Gaussian and Airy results to the exact value.}  
\label{tab:2}
\end{table}

\section{Discussion}
In this study, we show that AFM for a zero-dimensional 4-fermi model does work well as in the bosonic case\cite{rf:kashiwa} when taking higher ``loop" --- $1/N$ expansion in the Gaussian case and $1/N^{1/3}$ expansion in the Airy case --- effects into account even if $N$ is in the smallest number $N=2$.

So far we have assumed that $\lambda^2$ is real but of course it is not necessary so. However, there arises nothing new when we generalize $\lambda^2$ to be complex. In the complex $\lambda$ plane, there are two areas, located on the imaginary axis around $\lambda \approx \pm i/2$ (see 
eq.(\ref{s22})), where usual saddle point approximation breaks down; caustics. Other regions can be treated in a usual manner, that is, more accurate answer can be obtained when taking higher-loop effects into account.

Armed with this study, we should proceed to check the validity of AFM in quantum mechanical models. The work is now under progress\cite{rf:KS}. After this, if we would ascertain the validity of AFM further, we could draw a conclusion of the issue; whether the quantum effect of pions does wipe out the symmetry breaking\cite{rf:kleinert} or not\cite{rf:BLO}.

\vspace{20mm}

\noindent
{\Large \bf Acknowledgements}

\vspace{5mm}

\noindent 
This work supported in part by Grant-in-Aid for Science Research from the Japan Ministry of Education, Science and Culture: 12640280 and 13135217.

\end{document}